\documentclass[prd,aps,nofootinbib]{revtex4}
\usepackage{epsfig}
\usepackage{amsfonts}
\usepackage{rotating}
\usepackage{sparticles}
\usepackage{dcolumn}
\usepackage{bm}
\usepackage{hyperref}
\hypersetup{
    bookmarks=true,         % show bookmarks bar?
    unicode=false,          % non-Latin characters in Acrobat’s bookmarks
    pdftoolbar=true,        % show Acrobat’s toolbar?
    pdfmenubar=true,        % show Acrobat’s menu?
    pdffitwindow=false,     % window fit to page when opened
    pdfstartview={FitH},    % fits the width of the page to the window
    pdftitle={My title},    % title
    pdfauthor={Author},     % author
    pdfsubject={Subject},   % subject of the document
    pdfcreator={Creator},   % creator of the document
    pdfproducer={Producer}, % producer of the document
    pdfkeywords={keyword1} {key2} {key3}, % list of keywords
    pdfnewwindow=true,      % links in new window
    colorlinks=true,       % false: boxed links; true: colored links
    linkcolor=red,          % color of internal links
    citecolor=cyan,        % color of links to bibliography
    filecolor=magenta,      % color of file links
    urlcolor=green,           % color of external links
    linktocpage=true
}
\usepackage{amsmath,amssymb,amsthm,graphicx,latexsym}
\usepackage{subfigure}
\usepackage{pstricks}
\usepackage{color}

\usepackage{mathrsfs}

\newcommand{\be}{\begin{equation}}
\newcommand{\ee}{\end{equation}}
\newcommand{\bea}{\begin{eqnarray}}
\newcommand{\eea}{\end{eqnarray}}
\newcommand{\bml}{\begin{subequations}}
\newcommand{\eml}{\end{subequations}}
\newcommand{\bfig}{\begin{figure}}
\newcommand{\efig}{\end{figure}}

\newcommand{\slashed}{\hspace{-1.1ex}/}

\def\bea{\begin{eqnarray}}
\def\eea{\end{eqnarray}}
\def\ba{\begin{array}}
\def\ea{\end{array}}

\def\beq{\begin{equation}}
\def\eeq{\end{equation}}

\begin{document}

\title{\LARGE \textsc{\fontsize{95}{100}\selectfont \sffamily \bfseries{Quantum Gravity Effect in Torsion Driven Inflation and CP violation }}}

\author{\bf \textsc{\fontsize{10}{15}\selectfont \sffamily \bfseries{Sayantan Choudhury \footnote{\bf\textcolor{red}{\bf Electronic address: {sayantan@theory.tifr.res.in, sayanphysicsisi@gmail.com}}}}}}
\affiliation{\textsc{\fontsize{10}{15}\selectfont \sffamily \bfseries{Department of Theoretical Physics, Tata Institute of Fundamental Research, Colaba, Mumbai - 400005, India
\footnote{\textcolor{red}{\bf Presently working as a Visiting (Post-Doctoral) fellow at DTP, TIFR, Mumbai.}}}}}

\author{\bf \textsc{\fontsize{10}{15}\selectfont \sffamily \bfseries{Barun Kumar Pal \footnote{\bf\textcolor{blue}{\bf Electronic address: {barunp1985@rediffmail.com}}}}}}
\affiliation{\textsc{\fontsize{10}{15}\selectfont \sffamily \bfseries{Inter-University Centre for Astronomy and Astrophysics, Ganeshkhind, Pune 411007, India\\
Netaji Nagar College for Women, Regent Estate, Kolkata 700092, India}}}

\author{\bf \textsc{\fontsize{10}{15}\selectfont \sffamily \bfseries{Banasri Basu \footnote{\bf\textcolor{green}{\bf Electronic address: {sribbasu@gmail.com}}}}}}
\affiliation{\textsc{\fontsize{10}{15}\selectfont \sffamily \bfseries{Physics and Applied Mathematics Unit, Indian Statistical Institute, 203 B.T. Road, Kolkata 700 108, India}}}

\author{\bf \textsc{\fontsize{10}{15}\selectfont \sffamily \bfseries{Pratul Bandyopadhyay \footnote{\bf\textcolor{black}{\bf Electronic address: {b${}_{-}$pratul@yahoo.co.in}}}}}}
\affiliation{\textsc{\fontsize{10}{15}\selectfont \sffamily \bfseries{Physics and Applied Mathematics Unit, Indian Statistical Institute, 203 B.T. Road, Kolkata 700 108, India}}}

%\vspace{5ex}
%\date{\today}
\begin{abstract}
We have derived an effective potential for inflationary scenario from torsion and quantum gravity
correction in terms of the scalar field hidden in torsion. 
%In this work we propose an Effective Field Theory of inflation taking into account the spin density of matter which contributes to torsion.
 %%We first explicitly show that 
%torsion mimics the role of a scalar field which controls the dynamics of inflation. We have obtained a
 A strict bound on the CP violating $\theta$ parameter, ${\cal O}(10^{-10})<\theta<{\cal O}(10^{-9})$ has been obtained,
using {\tt Planck+WMAP9} best fit cosmological parameters.
\end{abstract}

%\pacs{98.80.-k ; 98.80.Cq ; 04.50.-h}

\maketitle
\section{\textsc{\fontsize{10}{15}\selectfont \sffamily \bfseries{Introduction}}}
The paradigm of cosmic inflation complements the big-bang theory and when combined together it is the best theory compatible 
with the latest observations. Inflation is generally believed to be driven by a scalar field known as inflaton. %, but there are other candidates as well.  
%The general features of inflationary scenario are believed to be caused by a scalar field called inflaton. Poplawski
Gasperini \cite{gasperini} has pointed out that the inflationary scenario can be well explained through torsion~\footnote{Also it is important to note that, in Ref.~\cite{Alexander:2009uu}, the authors 
have explicitly studied the late-time cosmic acceleration from torision and
the emergent scalar degree of freedom arose from the BCS condensation of
the fermions.}. Later, Poplawski \cite{pop1} have argued that torsion can be treated as an alternative source of inflation.
%In Ref. \cite{pop1},  it has been shown that the torsion can be treated as an alternative source of inflation. In fact, 
%inflationary scenario can also be well explained through torsion and there is a hidden scalar field in torsion were first pointed out in Ref. \cite{gasperini}. 
In this context, it is also to be noted that when torsion is considered to be generated from spin-spin interaction a hidden scalar field can be associated with torsion
\cite{mullick}. %In view of this, "the scalar field driven inflation"  as well as the "torsion driven inflation" appears to be equivalent statements.
%are equivalent statements.  
It is interesting to know whether 
the associated scalar field in torsion plays the role of inflaton in the inflationary regime such that "the scalar field driven inflation"  as well as the "torsion driven inflation" appear to be equivalent statements.
%are equivalent statements.  .  We found that, it is indeed the case when we deal
The motivation of the present paper is to show that if in the Einstein-Cartan-Kibble-Sciama (ECKS) theory  of gravity, 
the quantum gravity effect in the early universe is taken into account, we can formulate an effective potential for inflation in terms of the scalar field hidden in the torsion. Besides, the formulation gives rise to a CP violating term.  % Taking resort to this quantum correction we can formulate an effective potential generating the inflation. 
The estimate of the bound on CP violation in the early Universe  using {\tt Planck+WMAP9} best fit cosmological parameters \cite{Ade:2013uln} has also been obtained. 
\section{\textsc{\fontsize{10}{15}\selectfont \sffamily \bfseries{Torsion induced potential}}}
To study torsion in terms of the spin-spin interaction we take resort to a spin-current duality relation so that the action for torsion can be developed 
through a dual current-current interaction. We consider a four vector $n_{\mu}$ in terms of the spinorial variables as  
 %At the classical level a generalized theory of gravity incorporating torsion is given by the celebrated Einstien-Cartan-Kibble-Sciama (ECKS) formalism \cite{tom}~\footnote{The ECKS theory \cite{tom} has two fold advantages over GR because-(1) torsion appears to prevent the formation of singularities
%from matter composed of particles with half-integer spin and averaged as a spin fluid, and (2) to introduce an
% UV cutoff in quantum field theory for fermionic degrees of freedom.}.
%In this, the affine connection has non-vanishing antisymmetric contribution leading to torsion which can be represented by spin-spin interaction. 
%A dual current-current interaction picture can be developed  
%by translating
%the ${\bf SU(2)}$ spin basis into the topological current. 
%Within this prescription the spin-current duality can be explained in terms of a four-vector $n_{\mu}$ in the Casimir operator basis as:
\begin{equation}
n_\mu=\left(\frac{1}{\sqrt{2}}\right)(\psi_1^*~~~\psi_2^*) \sigma_\mu \left( \begin{array}{c} \psi_1 \\
                                                             \psi_2
                                                             \end{array}\right)
                                                             \end{equation}
where \be \psi_1 = (\cos\theta/2)e^{i\phi/2},~~~
\psi_2 = (\sin\theta/2)e^{-i\phi/2},\ee with $\sigma_0 =I$, where $I$ is the identity matrix and $\vec{\sigma}$ is the vector of Pauli matrices. Using this one can construct an
{\bf SU(2)} group element \be g=n_0I+i~\vec{\bf n}.\vec{\bf{\sigma}},\ee in terms of which we can construct the topological current  
 as \cite{Abanov:1999qz}:
\beq\label{kac}
J_\mu=\left( \frac{1}{24\pi^2}\right) \epsilon_{\mu\nu\lambda\sigma}{\rm Tr}[ (g^{-1}\partial^\nu g)(g^{-1}\partial^\lambda g)(g^{-1}\partial^\sigma g)]
\eeq
where $\epsilon_{\mu\nu\lambda\sigma}$ is the rank-4 Levi-Civita tensor.
Now by demanding that in 4-dimensional Euclidean  space the field strength $F_{\mu\nu}$ of a gauge potential vanishes
on the boundary $S^3$ of a certain volume ${\rm Vol_4}$ inside of which  $F_{\mu\nu} \neq 0$, we can write the gauge potential
 as %\beq\label{eqa}
$A_\mu=g^{-1}\partial_\mu g\in {\bf SU(2)}$.
%\eeq
Then from Eq.(\ref{kac}) the Kac-Moody like current $J_\mu$ can be recast in terms of the
 Chern-Simons secondary characteristic class as \cite{cs}:
\beq\label{ed1}
J_\mu=\left( \frac{1}{16\pi^2}\right) \epsilon^{\mu\nu\lambda\sigma}{\rm Tr} \left(A_\nu F_{\lambda\sigma}+\frac{2}{3}A_\nu A_\lambda A_\sigma \right)
\eeq
This gives rise to a topological invariant :
\beq\label{ponty}
Q_{P}=\left( \frac{1}{16\pi^2}\right)\int d^{4}x ~\partial_{\mu}J^{\mu}
\eeq
which is  known as the {\tt Pontryagin index}. %~\footnote{In fact the {\tt Pontryagin index} is related to the Chern-Simons invariant through the relation:
%$$\int_{{\cal M}_{4}}Tr(F\wedge F)=\int_{{\cal M}_{3}}Tr(A\wedge dA+\frac{2}{3}A\wedge A\wedge A)$$
%where $F$ is the two-form related to the field strength.}. 
%~This {\tt Pontryagin index} essentially corresponds to the action arising from torsion. 
We can construct the Lagrangian from the divergence of the current $J_{\mu}$
and write 
\beq
{\cal L}=-\frac{1}{4}{\rm Tr}\left( \epsilon_{\mu\nu\lambda\sigma}F^{\mu\nu}F^{\lambda\sigma}\right)
\eeq
 which 
leads to the construction of the current \cite{carmeli}
\beq
{\bf{j}}^\mu=\epsilon^{\mu\nu\lambda\sigma} {\bf a}_\nu \otimes {\bf f}_{\lambda\sigma}=\epsilon^{\mu\nu\lambda\sigma}\partial_\nu {\bf f}_{\lambda\sigma} 
\eeq 
with $A_\mu={\bf a}_\mu.{\bf{\sigma}}$
and %following \cite{carmeli}
\beq F_{\mu\nu}=\partial_{[\mu}A_{\nu]}+[A_{\mu},A_{\nu}]={\bf f}_{\mu\nu}.{\bf{\sigma}} \eeq
 It can be shown that the axial vector current 
\be J_{\mu}^5=\bar{\psi}\gamma_\mu\gamma_5\psi \ee 
is related to the second component of the current ${\bf j}_{\mu}$ through the relation \be \partial^\mu j_\mu^{(2)}=-\frac{1}{2}\partial^\mu J_\mu^5\neq0.\ee 
The consistency of the current conservation equations implies that \cite{aroy}:
\beq
j_\mu^{(1)}=-\frac{1}{2}j_\mu^{(2)},j_\mu^{(3)}=+\frac{1}{2}j_\mu^{(2)}
\eeq
Consequently, the current-current interaction can be expressed in terms of $j_\mu^{(2)}$ only which effectively displays the spin-spin interaction. Now we can write
%Indeed, the torsion arising from the current-current interaction of fermions is then given by the topological action as:
%\begin{eqnarray}
%S_{T}&=&\Sigma\int d^{4}x~ J_{\mu}J^{\mu}= 
%\end{eqnarray}
the action for torsion as \cite{aban}
\beq
S_T=\frac{M^{2}_{p}}{2} \int J_\mu^2 J_\mu^2 d^4x
\eeq
where $M_p$ being the reduced Planck mass, given by $M_p \approx 2.43\times 10^{18}$~GeV.         
It is now observed that there is a hidden scalar field $\phi$ in torsion which follows from the relation 
%Here we define a duality condition for the component $j_\mu^{(2)}$ as:
\begin{equation}
 {j}^{\mu (2)}=\epsilon^{\mu\nu\lambda\sigma}\partial_\nu {f}_{\lambda\sigma}^{(2)}
=\epsilon^{\mu\nu\lambda\sigma}{\epsilon}_{\nu\lambda\sigma}\phi(x)
\end{equation}
%where $\phi(x)$ is  the hidden scalar field and we will later see that it plays the role of inflaton, 
where ${\epsilon}^{\nu\lambda\sigma}$ is the rank-3 Levi-Civita tensor. 
The action now turns out to be:
\begin{eqnarray}\label{wqe1}
\small S_{T}%&=& %-\frac{\Sigma}{4}\int d^{4}x~ \epsilon_{\mu\nu\lambda\sigma}\epsilon^{\mu\alpha\beta\rho}~{\epsilon}^{\nu\lambda\sigma}\epsilon_{\alpha\beta\rho}~\phi^{2},\nonumber\\
 %&=& \frac{1}{2}\int d^{4}x~e~m^{2}\phi^{2},\nonumber \\
= %-\frac{\Sigma}{4}
{\cal A}\int d^{4}x~ j_{\mu}^{(2)}j^{\mu~(2)}
=\int d^{4}x~\sqrt{-g_{(4)}}~\frac{m^{2}}{2}\phi^{2}
\end{eqnarray}
%which actually represents the CP conserving contribution from torsion. %In Eq.(\ref{wqe1}) 
%$\Sigma$ represents the current-current interaction strength. %Within EFT the effective mass of the scalar degrees of
 %freedom is defined as~\footnote{Though Eq.(\ref{masseffect}) implies time dependence on the mass, but within the 
%standard inflationary description from EFT the time dependence of the scale factor $a(t)$ follows the de-Sitter solution so that the time dependence of mass freezes.
%In more generalized EFT prescription this time dependence on mass can give rise to various interesting features and the non-trivial behavior can be studied
% via Renormalization Group Flow Equations. }:
%\begin{eqnarray}\label{masseffect}
%m^{2}&=&-\frac{\Sigma}{2e}\epsilon_{\mu\nu\lambda\sigma}\epsilon^{\mu\alpha\beta\rho}~{\epsilon}^{\nu\lambda\sigma}\epsilon_{\alpha\beta\rho}
     %&=&-\frac{3\Sigma}{e}~{\epsilon}^{\alpha\beta\rho}\epsilon_{\alpha\beta\rho},\nonumber\\
 %    =-\frac{144\Sigma}{a^{3}(t)}
%\end{eqnarray}
%where $a(t)$ is the scale factor in FRW space-time.
Eq.(\ref{wqe1}) suggests that the potential associated with torsion can be written as:
\be V_T(\phi)=-\frac{m^2}{2}\phi^2.\ee
The negative sign of the coupling constant $m^2$ actually corresponds to the self interaction, when orientation of 
all the spin degrees of freedom 
are along the same direction.

%%%%%%%%%%%%%%%%%%%%%%%%%%%%%%%%%%%%%%%%%%%%%%%%%%%%%%%%%%%%%%%%%%%%%%%%%%%%%%%%%%%%%%%%%%%%%%%%%%%%%%%%%%%%%%%%%%%%%%%%%%%%%%%%%%%%%%%%%%%%%%%%%%%%%%%%%%%%%%%%%
%%%%%%%%%%%%%%%%%%%%%%%%%%%%%%%%%%%%%%%%%%%%%%%%%%%%%%%%%%%%%%%%%%%%%%%%%%%%%%%%%%%%%%%%%%%%%%%%%%%%%%%%%%%%%%%%%%%%%%%%%%%%%%%%%%%%%%%%%%%%%%%%%%%%%%%%%%%%%%%%%
\section{\textsc{\fontsize{10}{15}\selectfont \sffamily \bfseries{ INFLATIONARY MODELING WITH THE CP VIOLATING TERM}}}
Now we analyse the contribution from quantum gravity. To this end we %find the contribution from quantum gravity, we 
utilize the  model of Capovilla, Jacobson and Drell (CJD) \cite{cjd}, where the action is given by \cite{cjd}:
\beq
S=\frac{1}{8} \int \eta (\Omega_{ij}\Omega_{ij}+a\Omega_{ii}\Omega_{jj})
\eeq
where \be \Omega_{ij}=\epsilon^{\alpha\beta\gamma\delta}F_{\alpha\beta i}F_{\gamma\delta j}\ee
with $ \alpha, \beta, \gamma, \delta$ as space time indices, $i,j$ the ${\bf SU(2)}$ group indices and $\eta$ is a scalar density. 
In Ref.~\cite{cjd} it has been shown that in 3+1 decomposition this action yields Ashteker action directly provided we have 
$a= -\frac{1}{2}$ and the determinant of the magnetic field ${B^i}_a$ is non zero and as such  the equivalence to the Einstein's theory is  established.  The equivalence to the Einstein's theory can also be shown when the space time metric is found to be given by
\begin{equation}
\sqrt{-g_{(4)}}g^{\alpha\beta}= - \left(\frac{2i}{3\eta}\right)\epsilon_{ijk}\epsilon^{\alpha\gamma\delta\rho}
\epsilon^{\beta\mu\nu\sigma} F_{\gamma\delta i} F_{\rho\sigma j} F_{\mu\nu k}
\end{equation}
The constraint that is obtained when the CJD action is varied with respect to the Lagrangian multiplier $\eta$ is actually the Hamiltonian constraint
\begin{equation}
\Psi=\Omega_{ij} \Omega_{i j} ~-~\frac{1}{2} \Omega_{i i} \Omega_{j j}=i(2\eta^2~det B)^{-1}~H
\end{equation}
This implies that $\Psi \approx 0$ and $H\approx 0$ are equivalent statements provided $det B\neq 0$.
The canonical transformation of {\bf SU(2)} gauge potential ($A_{a i}$) and the corresponding non-abelian fields ($E_i^a,B_i^a$):
\bea A_{a i}&\rightarrow& A_{a i}, \\ E_i^a &\rightarrow& E_i^a - \theta B_i^a \eea 
%the expression for the Hamiltonian constraint changes though other constraints remain unaltered.
% The above transformation
 gives rise to a CP-violating $\theta$ term in the CJD Lagrangian so that for $a=-1/2$
%~\footnote{In 3+1 dimensional decomposition, $a=-1/2$ corresponds to Astekar's action.} 
 the action now reads \cite{cjd,mullick,pratul}:
\beq\label{astker}
S_{C}=\frac{1}{8} \int\left[ \theta\Omega_{ii}~+~\eta \left(\Omega_{ij}\Omega_{ij}-\frac{1}{2}\Omega_{ii}\Omega_{jj}\right)\right].
\eeq
In the first term the parameter $\theta$ essentially corresponds to the measure of CP violation which contributes to torsion and the rest is curvature contribution.
Consequently Eq.(\ref{astker}) can be recast as:
\begin{eqnarray}\label{astker1}
S_{C}&=&-\frac{\theta}{4}Q_{P}+\eta\int d^{4}x ~\epsilon^{\alpha\beta\gamma\delta}\epsilon^{\lambda\rho\sigma\mu}
\epsilon_{\nu\alpha\beta}\epsilon_{\nu^{'}\lambda\rho}\epsilon_{\xi\nu\delta}\epsilon_{\xi^{'}\sigma\mu}%\nonumber\\
%&&~~~~~~~~~~
\int dx^{\nu}~\phi \int dx^{\nu^{'}}~\phi\int dx^{\xi}~\phi\int dx^{\xi^{'}}~\phi \nonumber\\
&&~~~~~~~~~~~~~~~~~~~~~~~~~~~~~~~~~~~~~~~~~~~~~~~~~~~~~~~~~-\frac{\eta}{2}\int d^{4}x~ \epsilon^{\mu\nu\lambda\sigma}\epsilon^{\alpha\beta\gamma\delta}\epsilon_{\nu\lambda\sigma}
\epsilon_{\beta\gamma\delta}~(\partial_{\mu}\phi)(\partial_{\alpha}\phi)
\end{eqnarray}
where 
$\int dx^{\nu}~\phi=\phi~[x^{\nu}]$,
and the symbol $[\cdots]$ signifies the boundary value of the coordinates in the affine parameter space.
 Now from Eq.(\ref{astker1}) we get~\footnote{Here we use the following spin-particle duality relations:$$\eta\epsilon^{\mu\nu\lambda\sigma}\epsilon^{\alpha\beta\gamma\delta}\epsilon_{\nu\lambda\sigma}
\epsilon_{\beta\gamma\delta}=-~\sqrt{-g_{(4)}}~g^{\mu\alpha}$$ $$\eta\epsilon^{\alpha\beta\gamma\delta}\epsilon^{\lambda\rho\sigma\mu}
\epsilon_{\nu\alpha\beta}\epsilon_{\nu^{'}\lambda\rho}\epsilon_{\xi\nu\delta}\epsilon_{\xi^{'}\sigma\mu} [x^{\nu}x^{\nu^{'}}x^{\xi}x^{\xi^{'}}]=-\frac{\lambda}{4}.$$}:
\begin{eqnarray}\label{astker12}
\small S_{C}
=-\frac{\theta}{4}Q_{P}
+\int d^{4}x\sqrt{-g_{(4)}}\left[\frac{g^{\mu\alpha}}{2}(\partial_{\mu}\phi)(\partial_{\alpha}\phi)-\frac{\lambda}{4}\phi^{4}\right].
\end{eqnarray}
It may be mentioned here that the first term on the right hand side incorporates the Pontryagin index given by Eq.\eqref{ponty} which is a topological term arising from a total divergence. This does not contribute classically but has the effect in the quantum mechanical formulation.

From Eq~(\ref{wqe1}) and Eq~(\ref{astker12}), we note  that the action for torsion (curvature) when expressed in terms of the $\phi$ field involves the term $\phi^2 (\phi^4)$.
This indicates that the anisotropies associated with the torsion are much suppressed in comparison to the contribution from curvature for large values of $\phi$. It is noted that the %
%From our above discussion it appears that the scalar field here arises from gravitational degrees of freedom and thus is not a fundamental scalar field. 
%However, it is 
%to be mentioned that in this formalism there is a hidden scalar field associated with torsion. 
the expression of curvature in terms of the scalar field arises when we use 
CJD Lagrangian. In this sense the scalar field does not arise from gravitation as such, but it originates from the torsional degrees of freedom associated with the 
spin density.   
 
Noting that the asymptotic constancy of torsion compensates the bare cosmological constant \cite{Baekler:1987jb}
we can define a small but non-vanishing cosmological constant in terms of the Pontryagin index as 
\beq
M_p^2\Lambda_{eff}=\frac{\theta}{8~\rm Vol_{4}}Q_{P}
\eeq
where $M_p$ coresponds to the Planck mass. We can define the vacuum energy $V_0$ through the relation 
\beq
V_0=3 H^{2}_{inf}\Lambda^{2}_{UV}=\Lambda_{eff}\Lambda^{2}_{UV}
\eeq
Here $\Lambda_{UV}$ signifies the UV cut-off scale of
 the proposed EFT theory~\footnote{Above the scale $\Lambda_{UV}$ it is necessarily required to introduce the higher order quantum 
corrections to the usual classical theory of gravity represented via Einstein-Hilbert term,
 as the role of these corrections are significant in trans-Planckian scale to make the theory UV complete \cite{Assassi:2013gxa}. However such quantum 
corrections are extremely hard to compute as it completely belongs to the hidden sector of the theory dominated by heavy fields \cite{Choudhury:2014sxa}.
 In the trans-Planckian regime the classical gravity sector is corrected
by incorporating the 
effect of higher derivative interactions appearing through the modifications to GR which plays significant role in this context \cite{Choudhury:2013yg,Biswas:2011ar}.
On the other hand in trans-Planckian regime
quantum corrections of matter fields and their interaction between various constituents modify
 the picture which are appearing through perturbative loop corrections \cite{Assassi:2012et}.}. 
Below $\Lambda_{UV}$ the effect of all quantum corrections are highly suppressed and the heavy fields from the hidden sector gets their VEV. 
Such VEV is one of the possible 
sources of vacuum energy correction in the spin-current dominated EFT picture which uplifts the scale of inflationary potential and the contributions of the VEV become significant upto a scale 
$\Lambda_{C}\leq \Lambda_{UV}$. But at very low scale, $\Lambda_{low}\ll\Lambda_{C}$, one can tune the vacuum energy correction, $V_0\approx 0$ for which
 the contributions of the VEV can be neglected \cite{Allahverdi:2006iq}. 
Such possibility is only significant when the contribution of the primordial gravity waves become negligibly small (see Eq.(\ref{scalepot})).
Thus the expression for the potential from CJD Lagrangian incorporating the CP violating $\theta$ term yields:
\be V_C (\phi)=V_0+\frac{\lambda}{4} \phi^4.\ee

%%%%%%%%%%%%%%%%%%%%%%%%%%%%%%%%%%%%%%%%%%%%%%%%%%%%%%%%%%%%%%%%%%%%%%%%%%%%%%%%%%%%%%%%%%%%%%%%%%%%%%%%%%%%%%%%%%%%%%%%%%%%%%%%%%%%%%%%%%%%%%%%%%%%%%%%%%%%%%%%
%%%%%%%%%%%%%%%%%%%%%%%%%%%%%%%%%%%%%%%%%%%%%%%%%%%%%%%%%%%%%%%%%%%%%%%%%%%%%%%%%%%%%%%%%%%%%%%%%%%%%%%%%%%%%%%%%%%%%%%%%%%%%%%%%%%%%%%%%%%%%%%%%%%%%%%%%%%%%%%%
\section{\textsc{\fontsize{10}{15}\selectfont \sffamily \bfseries{The effective potential}}}
Now in the background of a space-time manifold having Riemannian structure %let us consider a situation where the superspace has Riemann structure. In such a case 
the contribution to the conserved current can be expressed as:
\be J^{\mu~g}=\frac{1}{2}\epsilon^{\mu\nu\lambda\sigma}R_{\nu\lambda\sigma\delta}v^{\delta},\ee
where $v^{\delta}$ is an arbitrary vector and Riemann curvature tensor can be expressed as:
\bea
R_{\nu\lambda\sigma\delta}=\partial_{[\lambda}\omega_{\nu]\sigma\delta}+\omega^{\eta}_{\nu\sigma}\omega_{\lambda\eta\delta}
-\omega^{\xi}_{\lambda\sigma}\omega_{\nu\xi\delta}-e_{\sigma\nu}e_{\delta\lambda}.
\eea 
As a result the gravitational part of the action can be written in terms of gravitational current-current interaction in the Riemann space as:
\begin{eqnarray}\label{gravac}
\small S_{g}= -\frac{\Lambda^{2}_{UV}}{2}\int d^{4}x J_{\mu}^{g}J^{\mu~g}=\frac{\Lambda^{2}_{UV}}{2}\int d^{4}x \sqrt{-g_{4}}R
\end{eqnarray}

Now clubbing the contributions from Eqns.(\ref{wqe1},\ref{astker12},\ref{gravac}) the total action for the present
field theoretic setup, taking into account quantum gravity correction,  can finally be written as:
\begin{eqnarray}
\label{totalaction}
\displaystyle S=\int d^{4}x\sqrt{-g_{4}}\left[\frac{\Lambda^{2}_{UV}}{2}R+\frac{g^{\mu\alpha}}{2}(\partial_{\mu}\phi)(\partial_{\alpha}\phi)-V(\phi)\right]
\end{eqnarray}
such that the total effective potential is given by:
\beq
V(\phi)=V_T(\phi)+V_C(\phi)=V_0 -\frac{m^2}{2}\phi^2+\frac{\lambda}{4}\phi^4.
\eeq

%%%%%%%%%%%%%%%%%%%%%%%%%%%%%%%%%%%%%%%%%%%%%%%%%%%%%%%%%%%%%%%%%%%%%%%%%%%%%%%%%%%%%%%%%%%%%%%%%%%%%%%%%%%%%%%%%%%%%%%%%%%%%%%%%%%%%%%%%%%%%%%%%%%%%%%%%%%%%%%%%%%%%
\section{\textsc{\fontsize{10}{15}\selectfont \sffamily \bfseries{Estimate on the CP violation term}}}
 The effective potential is dominated by the vacuum energy correction term which determines the scale of inflation. 
To obtain the scale of inflation at $k_{*}\approx k_{cmb}$, we express $V_0$ in terms of inflationary observables as:
\begin{eqnarray}\label{scalepot}
  V^{1/4}_{*}\approx V^{1/4}_0=7.389\times10^{-3}\Lambda_{UV}\times\left(\frac{r}{0.1}\right)^{1/4}.
\end{eqnarray}
where $r$ is the tensor-to-scalar ratio defined as: $r=A_T/A_S$ with $(A_T,A_S)$ being the amplitudes of the power spectra for scalar ($S$) and
tensor ($T$) modes at $k=aH\approx k_{*}$.
%~\footnote{Here we use the following parametrization for the power spectrum at any arbitrary momentum scale in the code CLASS: 
%$$A_{S}(k)=A_{S}\left(\frac{k}{k_{*}}\right)^{n_{S}-1+\frac{\alpha_S}{2}\ln\left(\frac{k}{k_*}\right)},~~A_{T}(k)=A_{T}\left(\frac{k}{k_{*}}\right)^{n_{T}}.$$}.
The effective cosmological constant or equivalently the CP violating parameter $\theta$ can then be constrained as:
\begin{eqnarray}\label{topopara}
 \Lambda_{eff}=\frac{\theta}{8~Vol_{4}}Q_{P}=2.98\times10^{-9}\Lambda^{2}_{UV}\times\left(\frac{r}{0.1}\right).
\end{eqnarray}

In order to compare the theoretical predictions with the latest observations we use a numerical 
code CLASS \cite{class}.  In this code we can directly  input the shape of the potential along with the model parameters. 
Then for a given cosmological background the code provides the estimates for different CMB observables. 
In the code we set the momentum pivot at $k_{*}=0.05$ Mpc$^{-1}$ and used the {\tt Planck + WMAP9} best fit values: \be h=0.670,~~ 
\Omega_b=0.049, ~~\Omega_c=0.268, ~~\Omega_\Lambda=0.682\ee for background cosmological parameters.
In this work we scan the parameter space within the following window:
\begin{eqnarray}\label{modpara}
 2.501\times 10^{-9}~\Lambda^{4}_{UV}\leq V_0 \leq 2.589\times 10^{-9}~\Lambda^{4}_{UV},\nonumber\\
6\times 10^{-3}~\Lambda^{-2}_{UV}\leq m^2 /V_0 \leq 8\times 10^{-3}~\Lambda^{-2}_{UV},\nonumber\\
\lambda /V_0 \sim 10^{-6}~\Lambda^{-4}_{UV}.
\end{eqnarray}
As a result, the CMB observables are constrained within the following range:
\begin{eqnarray}\label{cosmopara}
 2.197\times 10^{-9}&\leq A_S \leq& 2.202\times 10^{-9},\nonumber\\
  0.957   &\leq n_S \leq& 0.962,\nonumber \\
      -1.08 \times 10^{-3}    &\leq \alpha_S \leq& -0.99\times 10^{-3},\nonumber \\ 
            0.055           &\leq r \leq & 0.057. 
\end{eqnarray}
%%%%%%%%%%%%%%%%%%%%%%FIGURES%%%%%%%%%%%%%%%%%%%%%%%%%%%%%%%%%%%%%
\begin{figure}[t]
{\centerline{\includegraphics[width=14.0cm, height=9cm] {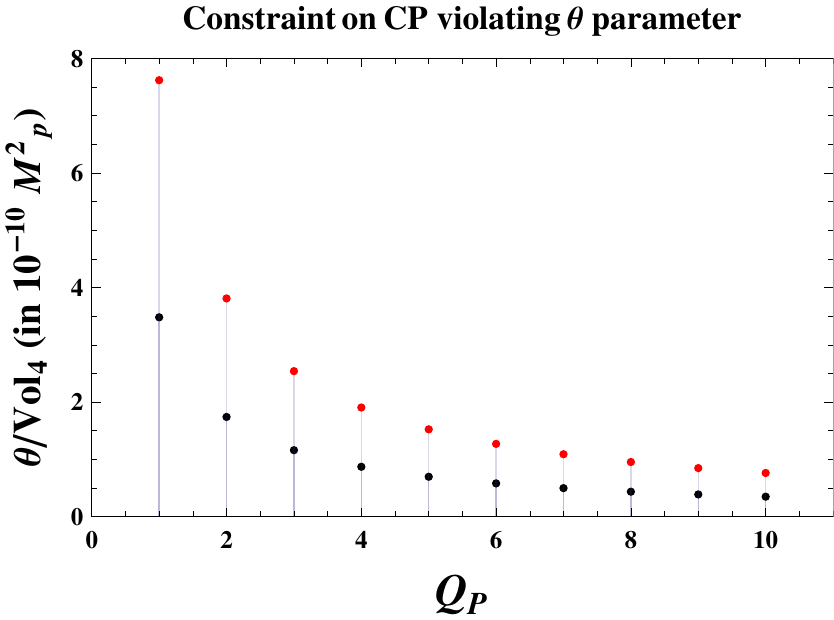}}}
\caption{Constraint on CP violating topological $\theta$ parameter for discrete integer values of {\it Pontryagin index} $Q_P$
using {\tt Planck + WMAP9} best fit cosmological parameters. Here \textcolor{red}{Red} and \textcolor{black}{\bf black} colored points correspond to the upper and lower bound of 
the $\theta$
parameter for a given value of $Q_P$. All the parallel \textcolor{blue}{blue} colored lines are drawn for different integer values of $Q_P$ which connects
both
the \textcolor{red}{Red} and \textcolor{black}{\bf black} colored points. This plot suggests that as the value of $Q_P$ increases then the interval between 
the upper and lower 
bound of the $\theta$ parameter decrease and it will converge to very small value for large $Q_P$. Also the numerical value corresponding to the upper bound and lower bound of the
$\theta$ parameter decreases once we increase the the value of $Q_P$. } \label{f2}
\end{figure}

%%%%%%%%%%%%%%%%%%%%%%%%%%%%%%%%%%%%%%%%%%%%%%%%%%%%%%%%%%%%%%%%%%%%%%%%%%%%%%%%%%

Within the present context the field excursion \cite{Lyth:1996im,Choudhury:2013iaa,Baumann:2011ws} is defined as:
\begin{eqnarray}
 \left|\Delta \phi\right|=\Lambda_{UV}\int^{N_{cmb}}_{0}dN\sqrt{\frac{r(N)}{8}}\approx \sqrt{\frac{r}{8}}N_{cmb}\Lambda_{UV}.
\end{eqnarray}
where $|\Delta \phi|=|\phi_{*}-\phi_{f}|$,
in which $\phi_{*}$ and $\phi_{f}$ represent the field value corresponding to CMB scale and end of inflation respectively. Also $N_{cmb}$ is
 the number of e-foldings at CMB scale which is fixed at $N_{cmb}\approx 50-70$
to solve the horizon problem associated with inflation. Subsequently we get the following constraint on the field excursion:
\be\left|\Delta \phi\right|\sim {\cal O}(4.1-5.9)\times\Lambda_{UV},\ee
which implies to make the EFT of inflation validate within the prescribed setup for which we need to constrain the UV cut-off of the EFT within the following window:
\be\Lambda_{UV}\sim {\cal O}(0.16-0.24)~M_p< M_p,\ee
which is just below the scale of reduced Planck mass. Finally using Eq.(\ref{topopara}) we get the following bound on the CP violating parameter
~\footnote{From experimental measurements of the neutron electric dipole moment, the experimental limit on the CP violating $\theta$ parameter is $\theta\leq 10^{-9}$ \cite{nair},
which is consistent with our derived stringent bound on $\theta$.}: 
\begin{eqnarray}\label{fb}
3.48\times 10^{-10}M^{2}_{p}\leq \frac{\theta}{Vol_{4}}Q_{P}\leq 7.62\times 10^{-10}M^{2}_{p}.
\end{eqnarray}

%%%%%%%%%%%%%%%%%%%%%%%%%%%%%%%%%%%%%%%%%%%%%%%%%%%%%%%%%%%%%%%%%%%%%%%%%%%%%%%%%%%%%%%%%%%%%%%%%%%%%%%%%%%%%%%%%%%%%%%%%%%%%%%%%%%%%%%%%%%%%%%%%%%%%%%%%%%%%%%%%
%%%%%%%%%%%%%%%%%%%%%%%%%%%%%%%%%%%%%%%%%%%%%%%%%%%%%%%%%%%%%%%%%%%%%%%%%%%%%%%%%%%%%%%%%%%%%%%%%%%%%%%%%%%%%%%%%%%%%%%%%%%%%%%%%%%%%%%%%%%%%%%%%%%%%%%%%%%%%%%%%
\section{\textsc{\fontsize{10}{15}\selectfont \sffamily \bfseries{Discussion}}}
Thus once we fix $Q_P$, this will further provide an estimate of $\theta$ according to the Eq.(\ref{fb}). In Fig.~(\ref{f2}) we
have explicitly shown the constraint on $\theta$ from the proposed EFT picture which is obtained by using {\tt Planck + WMAP9} best fit cosmological parameters.
To exemplify we have prescribed the 
bound on $\theta$ for different integer values of $Q_P$ lying within $1\leq Q_{P}\leq 10$.
From the plot it is easy to see that as the value of $Q_P$ increases the bound on the parameter 
$\theta$ converges to a very small value. This suggests that $\theta$ will converge to a constant value beyond a certain value of $Q_P$. It may be mentioned that the {\tt Pontryagin index}
can be taken to correspond to the fermion number \cite{Pratul1992}. Indeed a fermion can be realized as a scalar particle encircling a vortex line which is topologically
equivalent to a magnetic flux line and thus represents a {\tt skyrmion} \cite{Pratul1992}. The monopole charge $\mu=1/2$ corresponding to a magnetic flux line is related to the 
{\tt Pontryagin index} through the relation $Q_P =2\mu$. In view of this, one may note that $Q_P$ represents the fermion number which is the topological index carried 
by a fermion. For an anti-fermion $Q_P$ takes the negative value. In any system the effective fermion number is given by the difference between the number of fermions and 
anti-fermions. Thus we can quantify the fermionic matter and hence the spin density through the total accumulated value of $Q_P$. 
As $Q_P$ increases we have the increase of fermions implying the increase in spin density. So from Eq.(\ref{fb}) we note that for a fixed volume 
when $Q_P$ increases indicating the increase in spin density, the bound on the parameter $\theta$ converges to a small value representing the residual effect of torsion
residing at the boundary. Thus the remnant of CP violation~\footnote{In the context of canonical quantization of gravity it is observed that for small but non-vanishing value of the 
cosmological constant an exact solution to all the constraints of quantum gravity is given by the Chern-Simons state that describes the vacuum at the Planck scale which
 is chiral and implies an inherent CP-violation in quantum gravity \cite{kodama}.} giving rise to torsion can be witnessed through the small value of $\theta$ which is operative at the boundary.

%\section{summary}
 To summarize, we have derived an effective potential for inflationary scenario, taking into account the quantum gravity effect, in terms of the hidden scalar field associated with torsion along with a CP violating term.% and quantum gravity correction in terms of the scalar field. 
Using this %proposed a methodology for generating hidden scalar field within EFT framework from the spin spin interaction picture. We have explicitly computed 
%the vacuum energy corrected effective potential in sub-Planckian scale through which 
we give an estimate of inflationary CMB observables by constraining the model parameters-
vacuum energy, mass and self-coupling from {\tt Planck + WMAP9} best fit values of the cosmological parameters. 
Finally, for the first time we constrain the CP violating topological $\theta$ parameter from the %VEV of the
%heavy hidden sector fields appearing as 
vacuum  energy correction within EFT.

{\bf Acknowledgments:} 
 SC would like to thank Department of Theoretical Physics, Tata Institute of Fundamental
Research, Mumbai for providing me Visiting (Post-Doctoral) Research Fellowship. SC take this opportunity to thank
sincerely to Prof. Sandip P. Trivedi, Prof.
Shiraz Minwalla, Prof. Soumitra SenGupta, Prof. Sudhakar Panda, Prof. Varun Sahni, Prof. Sayan Kar, Prof. Sudhakar Panda, Dr. Subhabrata Majumdar and Dr. Supratik Pal for their constant support
and inspiration. SC take this opportunity to thank all the active members and the regular participants of weekly
student discussion meet “COSMOMEET” from Department of Theoretical Physics and Department of Astronomy
and Astrophysics, Tata Institute of Fundamental Research for their strong support. Last but not the least, we would
all like to acknowledge our debt to the people of India for their generous and steady support for research in natural
sciences, especially for theoretical high energy physics, string theory and cosmology.

%%%%%%%%%%%%%%%%%%%%%%%%%%%%%%%%%%%%%%%%%%%%%%%%%%%%%%%%%%%%%%%%%%%%%%%%%%%%%%%%%%%%%%%%%%%%%%%%%%%%%%%%%%%%%%%%%%%%%%%%%%%%%%%%%%%%%%%%%%%%%%%%%%%%%%%%%%%%%%%%%%%%%%%%%%%%%%%%%%%%%%%%%%%%%%%%%%%%%%%%%%%%%%%%%%%%%%%%%%%%%
%%%%%%%%%%%%%%%%%%%%%%%%%%%%%%%%%%%%%%%%%%%%%%%%%%%%%%%%%%%%%%%%%%%%%%%%%%%%%%%%%%%%%%%%%%%%%%%%%%%%%%%%%%%%%%%%%%%%%%%%%%%%%%%%%%%%%%%%%%%%%%%%%%%%%%%%%%%%%%%%%%%%%%%%%%%%%%%%%%%%%%%%%%%%%%%%%%%%%%%%

%\section*{Acknowledgments}

%%%%%%%%%%%%%%%%%%%%%%%%%%%%%%%%%%%%%%%%%%%%%%%%%%%%%%%%%%%%%%%%%%%%%%%%%%%%%%%%%%%%%%%%%%%%%%%%%%%%%%%%%%%%%%%%%%%%%%%%%%%%%%%%%%%%%%%%%%%%%%%%%%%%%%%%%%%%%%%%%%%%%%%%%%%%%%%%%%%%%%%%
%%%%%%%%%%%%%%%%%%%%%%%%%%%%%%%%%%%%%%%%%%%%%%%%%%%%%%%%%%%%%%%%%%%%%%%%%%%%%%%%%%%%%%%%%%%%%%%%%%%%%%%%%%%%%%%%%%%%%%%%%%%%%%%%%%%%%%%%%%%%%%%%%%%%%%%%%%%%%%%%%%%%%%%%%%%%%%%%%%%%%%%%%%%%%%%%%%%
%\section{Appendix} 

%%%%%%%%%%%%%%%%%%%%%%%%%%%%%%%%%%%%%%%%%%%%%%%%%%%%%%%%%%%%%%%%%%%%%%%%%%%%%%%%%%%%%%%%%%%%%%%%%%%%%%%%%%%%%%%%%%%%%%%%%%%%%%%%%%%%%%%%%%%%%%%%%%%%%%%%%%%%%%%%%%%%%%%%%%%%%%%%%%%%%%%%%%%
%%%%%%%%%%%%%%%%%%%%%%%%%%%%%%%%%%%%%%%%%%%%%%%%%%%%%%%%%%%%%%%%%%%%%%%%%%%%%%%%%%%%%%%%%%%%%%%%%%%%%%%%%%%%%%%%%%%%%%%%%%%%%%%%%%%%%%%%%%%%%%%%%%%%%%%%%%%%%%%%%%%%%%%%%%%

\end{document}